\documentclass[lettersize,journal]{IEEEtran}

\usepackage[OT1]{fontenc}
\usepackage[utf8]{inputenc}
\usepackage{amsmath,amssymb,graphicx,booktabs,tabularx,array}
\usepackage[expansion=false]{microtype}
\usepackage{balance}
\input{glyphtounicode}
\usepackage{cite}
\usepackage[hidelinks]{hyperref}
\hypersetup{pdftitle={SoulLink: Capability-Aware Continuous Affective Animation for 2D and 3D Avatars},pdfauthor={Yijun Feng; Zhijun Zhang; Zhi Chen; Jia Hua; Peilin Jin}}
\newcommand{\R}{\mathbb{R}}
\title{SoulLink: Capability-Aware Continuous Affective Animation for 2D and 3D Avatars}
\author{Yijun Feng, Zhijun Zhang, Zhi Chen, Jia Hua, and Peilin Jin%
\thanks{Yijun Feng and Zhijun Zhang are co-first authors. (Corresponding author: Peilin Jin.)}%
\thanks{Yijun Feng and Peilin Jin are with the College of Information Science and Technology, Shihezi University, Shihezi 832003, China. E-mail: \href{mailto:20241008398@stu.shzu.edu.cn}{\nolinkurl{20241008398@stu.shzu.edu.cn}}; \href{mailto:jinpeilin@shzu.edu.cn}{\nolinkurl{jinpeilin@shzu.edu.cn}}.}%
\thanks{Zhijun Zhang is with the College of Science, Shihezi University, Shihezi 832003, China. E-mail: \href{mailto:zhangzhijun20@xjshzu.com}{\nolinkurl{zhangzhijun20@xjshzu.com}}.}%
\thanks{Zhi Chen is with the Zhongfa Aviation Institute, Beihang University, Hangzhou 311115, China. E-mail: \href{mailto:24571137@buaa.edu.cn}{\nolinkurl{24571137@buaa.edu.cn}}.}%
\thanks{Jia Hua is with the School of Electronic Information, Wuhan University, Wuhan 430072, China. E-mail: \href{mailto:2024302121076@whu.edu.cn}{\nolinkurl{2024302121076@whu.edu.cn}}.}%
}

\begin{document}
\maketitle
\begin{abstract}
Real-time avatars express affect through heterogeneous controls that can be coupled, unavailable or shared with concurrent behaviors. SoulLink realizes continuous affect on existing Live2D and VRM rigs using a model-specific cue map and constrained temporal allocation. A quadratic program jointly accounts for reconstruction, continuity, activation ranges, rate limits and behavior-dependent resource budgets. Controlled cases characterize geometric coupling, capacity handoffs and response--smoothness tradeoffs. A block-randomized study with 386 participants, eight avatar conditions and 3,860 responses compares Full with static quadratic allocation followed by exponential smoothing. At 60 Hz and a matched 150 ms reference response, intended-emotion agreement increased from 87.80\% to 94.13\%, and recognition of both transition endpoints increased from 76.85\% to 88.49\%. Combined neutral--calm confusion decreased by 51.0\%. Change clarity improved by 0.190 points on a seven-point scale; the naturalness interval included zero. Solver diagnostics connect the coupled configuration's higher allocation cost to additional projected-gradient iterations and active resource projections. The results support joint temporal allocation for recognizable expression without a uniform reduction in reconstruction error or jerk.
\end{abstract}

\section{Introduction}
\IEEEPARstart{I}{nteractive} characters must express affect while speaking, blinking and looking. These behaviors share a limited set of controls: speech can reduce the mouth capacity available for a smile, while a new reaction can arrive before the previous expression has settled. A further challenge arises across embodiments. Live2D characters expose artist-defined parameters, whereas VRM avatars may combine individual morph targets and composite expressions. The same requested emotion therefore encounters different control capabilities and couplings on different characters.

Valence, arousal and dominance provide a compact description of affective input\cite{mehrabian1996}. FACS describes observable facial actions and supplies useful terminology for facial animation\cite{ekman2002}. Translating these representations to a heterogeneous rig requires an additional control model: which actuators exist, what their activation produces, and how their availability changes over time. Anatomical action-unit fidelity and preservation of a broader expressive impression are distinct objectives, particularly on stylized assets.

SoulLink formulates expression realization as allocation over an existing rig. Given an expressive target, a capability profile and a capacity schedule, it computes a feasible activation at each frame. A cue-effect map represents how direct and composite controls contribute to the shared expressive target. Temporal optimization then coordinates progress toward that target with the rig's available capacity and allowed motion. The implementation integrates the continuous affect machinery of Soullink Emotion SDK with the VRM application SoullinkEmotion3D; their public project repositories are listed in the code and data availability statement.

We compare joint temporal allocation with a strong static quadratic-programming baseline followed by exponential smoothing. Both use the same capability representation and resource constraints, and their main comparison matches the reference-step response at 150 ms. Our evaluation combines controlled numerical diagnostics with a $2\times8$ block-randomized perceptual study. The latter tests state recognition, recognition of emotion transitions and subjective motion quality, revealing improvements in subtle neutral--calm distinctions and joint transition recognition that are not summarized by proxy-space RMSE alone.

The contributions are a common capability-aware allocation model for heterogeneous Live2D and VRM controls; an implementation and benchmark that isolate temporal constraints, control coupling and behavior ownership; and a perceptual evaluation of 386 participants that links the allocation strategy to recognizable expression and clearer transitions. Public backend repositories, frozen source snapshots and reproducible analysis scripts support reuse of the system.

\section{Related work}
\subsection{Affect and multimodal behavior realization}
ALMA separates emotion, mood and personality in a layered computational affect model\cite{gebhard2005}. SoulLink similarly treats affect as an intermediate state, but focuses on its realization through the controls of a supplied avatar. Its VAD-to-expression rules are authored animation choices.

AsapRealizer provides an extensible architecture for behavior realization\cite{reidsma2013}, and Elckerlyc supports continuous, tightly coordinated multimodal interaction\cite{welbergen2010}. These systems establish the importance of timing and embodiment in virtual-human applications. Our formulation addresses the smaller allocation problem inside an expressive actuator set whose capacity can be reduced by another behavior owner. It complements higher-level behavior planning rather than replacing it.

\subsection{Expression transfer and stylized rigs}
Facial retargeting with automatic range-of-motion alignment transfers expression semantics between facial control spaces\cite{ribera2017}. Expression Packing derives compact reference expressions for blendshape transfer\cite{carrigan2020}. Both concern relationships between facial models and their expression spaces. SoulLink assumes that a binding is already available and models how to use its available actuators during execution.

AU-Blendshape introduces AU-oriented stylized expression data and a learned expression basis across identities\cite{li2025}. This is closely related to the expressive representation problem. Its learned geometry and expression-space construction differ from the present use of supplied two-dimensional parameter rigs and three-dimensional morph rigs. An empirical comparison with learned retargeting requires compatible assets and a shared task; the current implementation comparisons concern allocation over supplied controls and explicit capability masks.

\subsection{Speech-driven emotional 3D facial animation and disentanglement}
Recent speech-driven systems learn relationships between audio, expression and facial motion. EMOTE separates speech-related and emotional behavior using complementary perceptual supervision and content--emotion exchange\cite{danecek2023emote}. EmoTalk disentangles speech content and emotion and fuses them with identity information to synthesize expressive 3D motion\cite{peng2023emotalk}. FaceXHuBERT conditions speech-derived motion on identity and emotional expressiveness\cite{haque2023facexhubert}. These methods address how to generate an expressive performance from speech.

Learned motion representations also address ambiguity and temporal coherence. CodeTalker uses a discrete motion prior\cite{xing2023codetalker}, SelfTalk couples speech recognition and lip reading during training\cite{peng2023selftalk}, and KMTalk reconstructs a full motion sequence from key-motion embeddings\cite{xu2024kmtalk}. FaceDiffuser generates stochastic animation in both vertex and blendshape representations\cite{stan2023facediffuser}. ProbTalk3D adds emotion and intensity control to probabilistic codebook sampling\cite{wu2024probtalk3d}. EmoVOCA combines speech-related and expressive motion to construct emotional 3D training data\cite{nocentini2025emovoca}.

These output spaces are heterogeneous. Some systems learn motion in a dataset-specific mesh or parametric face representation; others predict rig coefficients. In particular, CSTalk models correlations between MetaHuman controls and uses them to supervise emotional animation\cite{liang2024cstalk}. Accordingly, topology dependence should be assessed per representation rather than assumed for every neural method. Learned correlation or smoothness losses also have a different role from explicit, per-frame feasibility constraints.

SoulLink operates at the downstream \emph{rig realization layer}. Once an input has been converted to its common cue signature, allocation enforces the supplied activation bounds, disjoint capacity budgets and rate allowances, with a recorded exception when abrupt hard ownership conflicts with continuity. Inputs can originate in authored rules or language-model emotion commands. Neural facial generators could likewise supply targets through an appropriate conversion; integrating and evaluating such a conversion remains an extension. The present comparisons isolate allocation over identical targets and do not claim a direct performance ranking against speech-to-motion generators.

\subsection{Control interfaces and perceptual assessment}
ARKit blendshape names, ICT-FaceKit correspondences and VRM expression specifications provide practical exchange conventions\cite{applearkit,ictfacekit,vrmexpr}. Live2D exposes numeric parameter operations on artist-defined rigs\cite{live2dparams}. The visible meaning and coupling of these controls remain asset-dependent. SoulLink therefore records control identities, ranges and cue effects separately from exchange-protocol naming.

Dimensional ratings provide complementary information to categorical recognition. Bradley and Lang's work motivates distinguishing valence, arousal and dominance judgments\cite{bradley1994}. The platform study analyzed here focuses on categorical recognition, naturalness and the perceived continuity and clarity of transitions. These endpoints evaluate interpretation of the rendered expression, whereas numerical tracking measures characterize the controller.

\section{System representation}
Let $\mathbf{e}_t\in[-1,1]^3$ denote a VAD input and let $\mathbf{z}_t\in[0,1]^k$ denote a desired expressive signature. The reference implementation uses $k=12$ cues covering brow, eye and mouth appearance and vertical head pose. Scripted cues or the existing SDK and 3D expression modules can produce this signature. Allocation comparisons use identical target samples for a given trial.

For avatar $m$, $\mathbf{q}_t\in\R^{n_m}$ contains normalized nonnegative control activations. A control can represent a single morph, a composite expression or one direction of a signed parameter. The binding layer converts activations to physical parameters, including neutral offsets and signed scales. A paired Live2D eye control can write both left and right parameter IDs. Opposite directions of a shared parameter use a common resource group.

\begin{figure}[!t]\centering
\includegraphics[width=\linewidth]{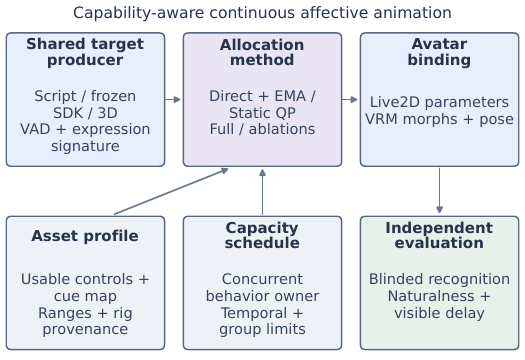}
\caption{SoulLink architecture. A shared target signature feeds all allocation methods. An asset profile and a behavior-capacity schedule constrain the controller, whose physical outputs drive the selected embodiment.}\label{fig:architecture}
\end{figure}

Figure~\ref{fig:architecture} separates target production, allocation and rendering. Upstream language and speech services are optional components. Scripted inputs isolate animation behavior from variability in text generation and network requests. The SDK and 3D application retain their own expression mappings; experiments comparing allocation use a shared target signature rather than treating a change of target producer as an allocation effect.

\section{Capability-aware allocation}
\subsection{Cue-effect model}
A profile records asset provenance, usable controls, physical bindings, upper bounds, rate limits, resource groups and a cue-effect matrix $C_m\in\R_{\ge0}^{k\times n_m}$. The visible signature is approximated locally as
\begin{equation}
\widehat{\mathbf{z}}_t=C_m\mathbf{q}_t.
\label{eq:cue}
\end{equation}
A column describes the cues associated with one actuator. Composite controls may have several nonzero entries, and multiple actuators may contribute to one cue. An unsupported control is omitted. A cue outside the span of the available columns retains a residual.

Profiles can be constructed using rating-slope fitting, range/alias templates or geometric decomposition. The rating protocol samples each control at levels $\lbrace0.25,\,0.50,\,0.75,\,1.00\rbrace$ under fixed framing. In rating-slope fitting, neutral-subtracted cue ratings are aggregated by their median at each level and fitted with a nonnegative through-origin slope; separate activation combinations assess additivity. The controlled independent-channel cases use range and alias templates. The coupled case estimates a composite column by nonnegative least-squares decomposition of position deltas in a regional-morph basis. These coefficients provide control-space proxies for the numerical diagnostics, while the perceptual experiment measures recognition of the rendered expression.

\subsubsection{Vertex-delta decomposition and profile inventory}
For the controlled VRM calibration, all morph directions use the asset's unchanged vertex correspondence. Let $\Delta V_{p,j}\in\mathbb{R}^{N_p\times3}$ denote the POSITION delta of regional morph $j$ on face primitive $p$, and $\Delta V_{p,c}$ the composite delta. Stacking vectorized deltas over the eight primitives gives
\begin{equation}
\begin{aligned}
B&=\bigl[\operatorname{stack}_p\operatorname{vec}(\Delta V_{p,j})\bigr]_{j=1}^{10},\\
\mathbf d_c&=\operatorname{stack}_p\operatorname{vec}(\Delta V_{p,c}),\\
\mathbf a_c&=\arg\min_{\mathbf a\ge0}\|B\mathbf a-\mathbf d_c\|_2^2.
\end{aligned}
\label{eq:geometry_nnls}
\end{equation}
The fit concatenates original POSITION deltas without per-column normalization or vertex-area weighting. It includes neither normals nor material changes. Sparse glTF accessors are expanded before stacking. The regional directions already share the composite's topology and coordinate system, so no registration or cross-mesh correspondence is introduced. The ten fitted coefficients populate the facial rows of the composite cue column; the two head-pose rows are zero. Thus $B$ models vertex displacement, whereas $C_m$ models the allocator's twelve-dimensional cue signature. Their spaces and units are distinct.

For \texttt{Fcl\_ALL\_Joy}, the normalized residual $\|B\mathbf a_c-\mathbf d_c\|_2/\|\mathbf d_c\|_2$ is 0.03438. Nonzero brow and eye coefficients accompany its mouth contribution, creating overlapping columns in $C_m$ and hence nonzero off-diagonal entries in $C_m^\top C_m$. This is a measured geometric coupling of supplied morphs, not an anatomical AU calibration or a perceptual intensity measurement.

Table~\ref{tab:calibration_profiles} reports the executable profiles used in the numerical cases. An actuator here is one normalized allocation variable; a signed parameter may require two directions, while one variable may write several physical IDs. Resource budgets therefore describe allocation groups rather than counts of raw SDK parameters or mesh blendshapes.
\begin{table*}[!t]\centering\small
\caption{Executable Controlled-Profile Inventory}
\label{tab:calibration_profiles}
\begin{tabularx}{\linewidth}{@{}p{.21\linewidth}rX p{.28\linewidth}@{}}\toprule
Profile & Actuators & Disjoint resource groups (count / budget) & Cue-map construction\\\midrule
Hiyori independent & 9 & Brow 3/1.0; eye 2/1.0; mouth 2/1.1; head 2/1.0 & Parameter-range template; unit proxy gains\\
VRM independent & 12 & Brow 3/1.0; eye 3/1.0; mouth 4/1.1; head 2/1.0 & Morph-name alias template; unit proxy gains\\
VRM composite mask & 12 & Face 10/1.0; head 2/1.0 & Ten regional directions; NNLS composite column; eight face primitives\\\bottomrule
\end{tabularx}
\par\smallskip
\begin{minipage}{\linewidth}\footnotesize
\textit{Note.} Group entries give actuator count / base budget. The independent profiles use range/alias templates; the coupled profile replaces an independent smile with a geometry-fitted composite. These three configurations describe the numerical diagnostics, not the eight anonymized perceptual condition codes.
\end{minipage}
\end{table*}

\subsection{Feasible controls and ownership}
Each activation has an upper bound $u_j$ and a rate limit $r_j$. The reference solver supports disjoint resource groups $G_g$ with time-dependent budgets $b_g(t)$:
\begin{align}
0\le q_{t,j}&\le u_j,\label{eq:box}\\
\sum_{j\in G_g}q_{t,j}&\le b_g(t),\label{eq:group}\\
|q_{t,j}-q_{t-1,j}|&\le r_j\Delta t.\label{eq:rate}
\end{align}
Resource groups describe the controls' joint activation limits. An ownership signal changes a group's available capacity, such as reducing the mouth budget when another behavior requires it. The owner reserves the mouth group on the independent profiles and the shared face envelope on the coupled profile. The reservation signal is prescribed; audio-driven lip synchronization is a separate integration task.

An abrupt budget reduction can conflict with Equation~\ref{eq:rate}. When the minimum activation reachable within a frame exceeds the new budget, the solver prioritizes the hard budget and relaxes the lower rate bound in that group. It records the resulting exception. This expresses a tradeoff between immediate ownership and continuous motion.

For a scheduled handoff, an owner announces the change with lead time $T>0$. Let $a\in[0,1]$ be the maximum fraction of the group's capacity reserved by that owner. We use a clamped smoothstep ramp:
\begin{equation}
\begin{aligned}
\rho(t) &= a\,s\!\left(\frac{t-t_{\mathrm{on}}+T}{T}\right)
\left[1-s\!\left(\frac{t-t_{\mathrm{off}}}{T}\right)\right],\\
s(v) &= 3\bar v^2-2\bar v^3,\qquad
\bar v=\operatorname{clip}(v,0,1).
\end{aligned}
\label{eq:ramp}
\end{equation}
The mouth budget is $b_{\mathrm{mouth}}(t)=b^0_{\mathrm{mouth}}[1-\rho(t)]$. The reserved fraction reaches $a$ at $t_{\mathrm{on}}$ and is released over $T$ after $t_{\mathrm{off}}$. An unannounced, immediate owner uses the hard-preemption policy above.

\subsection{Temporal objective and solution}
The full controller solves
\begin{equation}
\begin{aligned}
\min_{\mathbf{q}_t}\quad &
\frac{1}{2}\|C_m\mathbf{q}_t-\mathbf{z}_t\|_2^2
+\frac{\lambda}{2}\|\mathbf{q}_t\|_2^2\\
&+\frac{\eta}{2\Delta t}\|\mathbf{q}_t-\mathbf{q}_{t-1}\|_2^2.
\end{aligned}
\label{eq:objective}
\end{equation}
subject to Equations~\ref{eq:box}--\ref{eq:rate} and the preemption rule. The ridge term regularizes activation. The temporal term penalizes frame-to-frame change, with its coefficient scaled by the frame interval. With $\lambda>0$, the quadratic objective is strictly convex on a nonempty feasible set.

Warm-started projected gradient descent uses step size
$[\|C_m^\top C_m\|_2+\lambda+\eta/\Delta t]^{-1}$.
For each disjoint group, projection clips to the activation box and uses a scalar threshold to satisfy its budget. The implementation permits 60 gradient iterations, uses a residual tolerance of $10^{-7}$, and uses 45 bisection steps when a budget binds. Iteration counts and residuals are logged; Section~\ref{sec:solver_cost} analyzes convergence and projection work. Arbitrary overlapping groups require a more general projection or constrained solver.

\section{Implementation and comparisons}
The allocation reference uses Python and NumPy. TypeScript exporters execute frozen SDK commit \texttt{02cd75ac0a08} and Emotion3D commit \texttt{6697822e70d5}, recording the original states and cue conversions. Both snapshots retain their Apache-2.0 notices. A browser tool applies recorded activations to VRM morphs and head pose or to Live2D Core parameters. The library of source hashes and scripts accompanies the manuscript.

Table~\ref{tab:baselines} defines five implemented controllers.
\begin{table}[!t]\centering\small
\caption{Implemented Allocation Comparisons}\label{tab:baselines}
\begin{tabularx}{\linewidth}{@{}>{\raggedright\arraybackslash}p{.32\linewidth}>{\raggedright\arraybackslash}X@{}}
\toprule Method & Definition\\\midrule
Direct + EMA & Designated direct cue mapping, feasible projection, exponential smoothing and reprojection.\\
Static QP + EMA & Constrained cue least squares with ridge; exponential smoothing and reprojection.\\
Full & Equation~\ref{eq:objective}, rate limits and logged hard-preemption policy.\\
No temporal & Static constrained QP with ridge and no temporal filtering or rate limits.\\
\mbox{No reallocation} & Temporally regularized tracking of the direct mapped activation, with rate limits.\\\bottomrule
\end{tabularx}\par\smallskip
\begin{minipage}{\linewidth}\footnotesize
\textit{Note.} All methods use identical targets, activation bounds and group budgets.
\end{minipage}
\end{table}

 Static QP plus exponential smoothing is the principal comparator: it shares the cue map and hard resource constraints with Full. The initial five-method diagnostic uses $\tau=0.12$ s, $\eta=0.025$ s and $\lambda=0.002$. Reference-delay calibration selects $\tau=0.070$ s for comparison with Full at $\eta=0.025$ s. The perceptual stimuli use this calibrated setting at 60 Hz: Full uses $\eta=0.025$ s and $\lambda=0.002$, while Static QP + EMA uses $\tau=0.070$ s. Both share a reference-step response of $T_{90}=150$ ms.

Correctness checks compare small projections and quadratic solutions with an independent optimizer and exercise feasibility, frame-rate behavior and ownership exceptions. Target-machine checks also execute the original source exporters. Controller timing covers Python CPU allocation; browser application timing and network delay are separate measurements. The current implementation uses offline allocation followed by playback rather than a synchronized production TypeScript pipeline.

\section{Controlled implementation cases}
\subsection{Application preset repertoire}
\begin{figure*}[!t]\centering
\includegraphics[width=\linewidth]{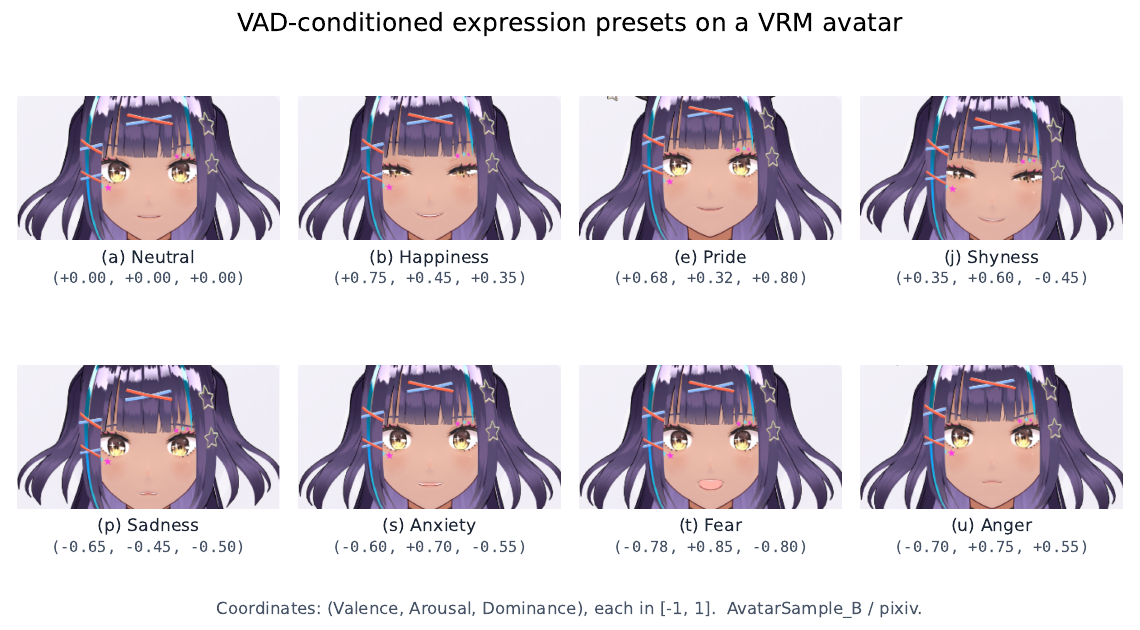}
\caption{Selected authored VRM presets on AvatarSample\_B. Triples indicate (valence, arousal, dominance), each in $[-1,1]$. The full 24-state atlas is supplied as supporting material. Character attribution: VRoid Project / pixiv.}\label{fig:atlas}
\end{figure*}

Figure~\ref{fig:atlas} shows eight representative states from a 24-preset VRM atlas supplied with the project. Each panel retains its authored label and VAD coordinates. The atlas uses expression gain 4, regional gains 2 and style-detail mix 1 to document the application's preset repertoire.

\subsection{Controller diagnostics}
Table~\ref{tab:quantitative} compares every method defined in Table~\ref{tab:baselines}. \begin{table*}[!t]\centering\small
\caption{Five-Method Controller Diagnostics}\label{tab:quantitative}
\setlength{\tabcolsep}{4pt}
\renewcommand{\arraystretch}{1.15}
\begin{tabularx}{\linewidth}{@{}Xrrrrrr@{}}\toprule
Method & $E$ & $V$ & $J/10^3$ & $T_{90}$ & CPU$_{50}$ & CPU$_{95}$\\\midrule
\multicolumn{7}{@{}l@{}}{\textit{Hiyori: independent alias profile}}\\[3pt]
Direct + EMA & 0.1496 & 0.295 & 1.193 & 266.7 & 0.061 & 0.130\\
Static QP + EMA & 0.1496 & 0.295 & 1.190 & 266.7 & 0.131 & 0.255\\
Full & 0.1489 & 0.492 & 3.636 & 150.0 & 0.115 & 0.222\\
No temporal & 0.1457 & 1.469 & 15.491 & 0.0 & 0.106 & 0.201\\
No reallocation & 0.1489 & 0.492 & 3.636 & 150.0 & 0.070 & 0.151\\
\midrule
\multicolumn{7}{@{}l@{}}{\textit{AvatarSample\_B: independent alias profile}}\\[3pt]
Direct + EMA & 0.0725 & 0.454 & 2.714 & 266.7 & 0.059 & 0.228\\
Static QP + EMA & 0.0725 & 0.453 & 2.706 & 266.7 & 0.136 & 0.471\\
Full & 0.0706 & 0.631 & 4.604 & 150.0 & 0.118 & 0.449\\
No temporal & 0.0590 & 1.766 & 18.088 & 0.0 & 0.107 & 0.433\\
No reallocation & 0.0706 & 0.631 & 4.603 & 150.0 & 0.075 & 0.249\\
\midrule
\multicolumn{7}{@{}l@{}}{\textit{AvatarSample\_B: coupled composite and capability mask}}\\[3pt]
Direct + EMA & 0.1244 & 0.487 & 3.301 & 266.7 & 0.145 & 0.265\\
Static QP + EMA & 0.1114 & 0.385 & 2.453 & 266.7 & 0.645 & 5.697\\
Full & 0.1085 & 0.521 & 3.379 & 150.0 & 0.291 & 3.276\\
No temporal & 0.1044 & 1.214 & 11.819 & 0.0 & 0.270 & 5.823\\
No reallocation & 0.1240 & 0.596 & 3.921 & 150.0 & 0.150 & 0.280\\
\bottomrule
\end{tabularx}
\par\smallskip
\begin{minipage}{\linewidth}\footnotesize
\textit{Note.} Measured on the target Windows machine (Ryzen 7 9800X3D, Python/NumPy CPU). Each configuration uses the same four scripted sequences at 60 Hz, five seconds each, with three timed repetitions. $E$: cue RMSE; $V$: RMS cue velocity (s$^{-1}$); $J$: RMS cue jerk (s$^{-3}$); $T_{90}$: absolute common-step response (ms); CPU$_{50/95}$: per-frame median/95th-percentile controller time (ms). Errors and derivatives are pooled over sequences; step delay uses the common feasible step. These are profile-space diagnostics.
\end{minipage}
\end{table*}

The configurations are the Hiyori Live2D alias profile, the independent AvatarSample\_B VRM alias profile, and the controlled composite configuration below. Each receives four five-second sequences at 60 Hz: a common feasible step, a target interruption, seeded target jitter and a prescribed ownership reservation. The ownership sequence reserves 65\% of the configured group from 1.50 to 4.50 s. The group is the mouth on the independent profiles and the conservative face envelope on the coupled profile. The same target samples, activation bounds, resource budgets and documented default hyperparameters are used within a configuration; temporal constraints differ as defined in Table~\ref{tab:baselines}.

Three timed repetitions use a fixed randomized method and task order. Each controller has an eight-frame, separate zero-input warm-up and starts the measured trial from neutral. Cue error and velocity are RMS values over frames and the twelve coordinates. Jerk is the RMS third finite difference of the reconstructed signature, scaled by the cube of the frame rate, computed separately within each sequence. The step uses brow-down 0.7 and head-up 0.2 from $t=1$ s; $T_{90}$ is the first frame attaining 90\% of the absolute requested vector projection. Its time resolution is 16.7 ms. CPU timing wraps the controller call with a monotonic high-resolution timer and excludes target generation, file I/O, network and rendering.

The full controller reaches the common step in 150 ms; the two EMA methods reach it in 266.7 ms. The no-temporal ablation responds in the first requested frame. With the coupled profile, the full controller's pooled cue RMSE is 0.1085, compared with 0.1114 for Static QP + EMA and 0.1244 for Direct + EMA. Its pooled jerk is 3379 s$^{-3}$, above Static QP + EMA (2453) and below No temporal (11819). These default settings therefore expose a response--smoothness tradeoff.\par
Full-controller CPU time at the 95th percentile ranges from 0.222 to 3.276 ms across the three configurations. Timing covers allocation alone on the stated machine. Hard range and resource violations remain below $10^{-8}$ for every method. The independent-profile Full and No reallocation results are close or identical; the coupled configuration separates their allocation behavior.

The table characterizes the alias and geometry-derived cue maps through deterministic controller trajectories. The initial EMA and Full settings represent different operating points. The reference-delay calibration below matches their common-step response before testing separate dynamic inputs. Recognition and perceived motion quality are evaluated independently in the participant study.

\subsection{Coupled expression controls and a capability mask}
The executable profile inventory in Table~\ref{tab:calibration_profiles} separates the controlled configurations. The coupled rendered case uses the VRM's existing \texttt{Fcl\_ALL\_Joy} composite, which changes brow, eye and mouth geometry together. Its position deltas are decomposed by nonnegative least squares over ten existing regional morph directions and all eight face primitives. The relative Euclidean position residual is 0.0344. The leading coefficients are 0.5445 for the surprised-brow direction, 1.0000 for eye joy and 1.0000 for mouth joy. This estimates a geometric relationship between supplied controls, not the perceived intensity of an emotion.

The mouth-smile coordinate in this case refers to \texttt{Fcl\_MTH\_Joy}; the independent alias configuration retains its original \texttt{Fcl\_MTH\_Fun} binding. Diagnostic values are compared between methods within a supplied profile.

\begin{figure*}[!t]\centering
\includegraphics[width=\linewidth]{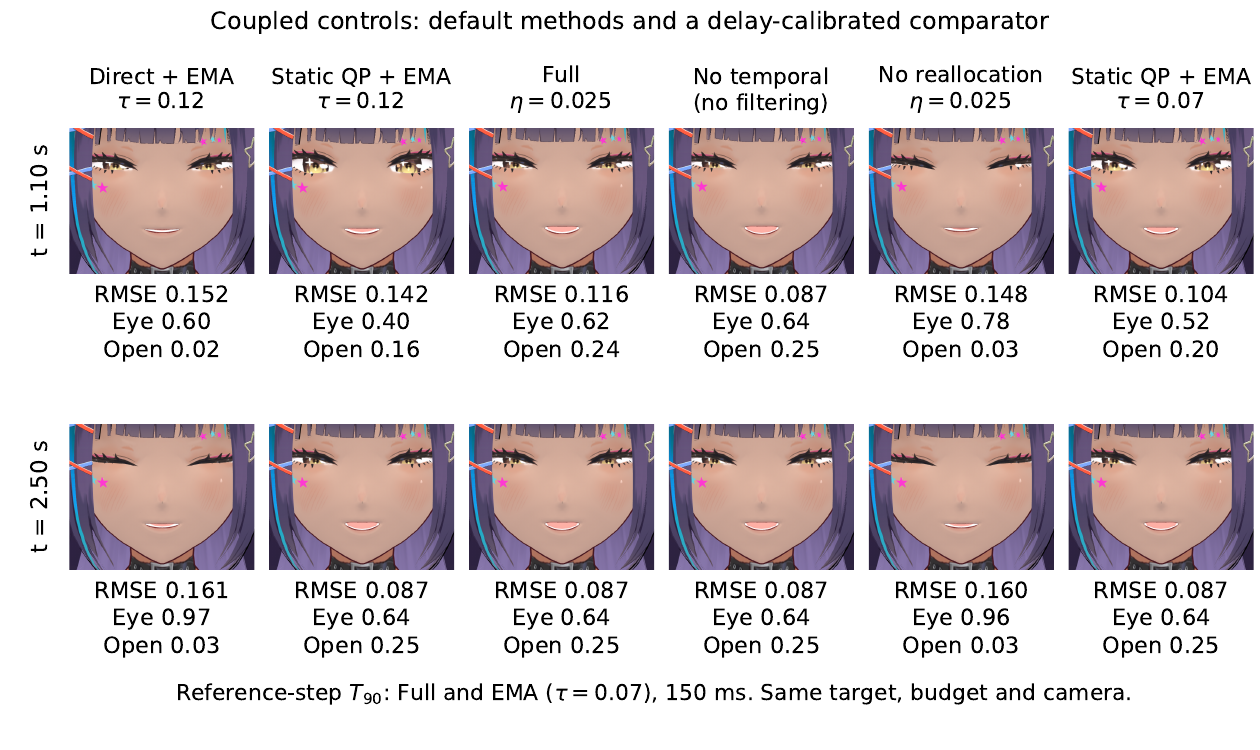}
\caption{Five default controllers and a reference-delay-calibrated static comparator with an unavailable independent smile actuator and an available coupled expression. The last column uses $\tau=0.070$ s; Full and that setting share a 150 ms reference-step response (the matched setting). Rows show samples at $t=1.10$ and $2.50$ s; the target begins at 1.00 s. Eye and Open labels are regional-morph proxy coefficients; the requested values are 0.55 and 0.25. The face budget and crop are identical across methods. The strong static comparator and Full approach the same steady allocation; the temporal ablations differ in their transition behavior. Character: AvatarSample\_B, VRoid Project / pixiv.}\label{fig:qualitative}
\end{figure*}

The independent \texttt{Fcl\_MTH\_Joy} actuator is omitted from the available controller set while the original asset remains intact. The composite remains available as the designated direct smile control. All facial actuators share one disjoint budget of 1.0; head controls retain their separate budget. This conservative envelope bounds the total activation of the composite and the remaining regional controls. The requested joy signature begins at 1.00 s. Figure~\ref{fig:qualitative} shows all five default methods and the reference-delay-calibrated static comparator 100 ms and 1.50 s after that onset, using identical close-up framing. Its annotations report instantaneous proxy RMSE, eye-smile reconstruction and mouth opening.

At time $t = 2.50$~s, Direct + EMA reconstructs an eye-smile coefficient of 0.967 and a mouth-opening coefficient of 0.033; Full produces 0.645 and 0.250, respectively. Static QP + EMA approaches the same steady allocation as Full, while No temporal reaches it immediately. The direct methods retain excess eye closure and suppress the independent mouth-opening control under the shared face budget. These visible differences follow from the coupled binding and the allocation objective.

The final column adds the calibrated comparator to the same rendered case. At 100 ms after onset, its proxy RMSE is 0.104, compared with 0.116 for Full and 0.142 for the default static comparator. Both calibrated and default static settings approach the same steady allocation. The reference-step match changes the interpretation of the transient visual comparison.

Additional fixed-camera replays on Hiyori and AvatarSample\_B are supplied as supporting material.

\subsection{Abrupt and announced capacity handoff}
A separate case holds a joy signature on the independent VRM profile while a concurrent owner reserves 80\% of mouth-group capacity from 1.50 to 2.50 s. One condition applies an abrupt reservation. The other uses Equation~\ref{eq:ramp} with $a=0.80$ and $T=0.35$ s, beginning at 1.15 s and reaching the same reserved capacity by 1.50 s. Both conditions use the same Full controller, profile and sampling rate.

The abrupt condition triggers one rate-exception frame, with a maximum normalized channel change of 0.6283 after the first second. The announced condition triggers no exception and has a maximum change of 0.0605 in that interval (Table~\ref{tab:preemption}). \begin{table}[!t]\centering\small
\caption{Scripted Capacity Handoff}\label{tab:preemption}
\begin{tabularx}{\linewidth}{@{}Xrrr@{}}\toprule Schedule & Exceptions & Max. $\Delta q$ & Rate excess\\\midrule
Hard reservation & 1 & 0.6283 & 0.5616\\
Announced ramp & 0 & 0.0605 & 0\\
\bottomrule\end{tabularx}
\par\smallskip
\begin{minipage}{\linewidth}\footnotesize
\textit{Note.} The AvatarSample\_B independent alias profile is sampled at 60 Hz. Both conditions use the same Full controller. The announced case has 350 ms look-ahead. Change and rate excess are measured in normalized activation units; zero excess is within numerical precision.
\end{minipage}
\end{table}

\begin{figure}[!t]\centering
\includegraphics[width=\linewidth]{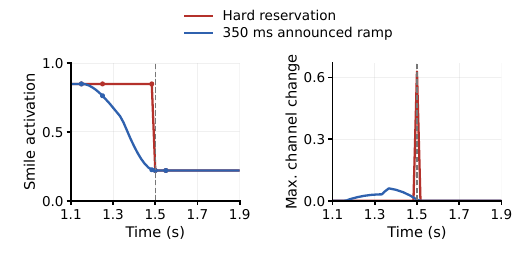}
\caption{Controller traces for the prescribed mouth-capacity handoff. The dashed line marks the 1.50 s deadline; points in the left plot mark the five rendered times in Figure~\ref{fig:handoff_frames}. The ramp receives 350 ms advance notice.}\label{fig:handoff_trace}
\end{figure}

\begin{figure*}[!t]\centering
\includegraphics[width=\linewidth]{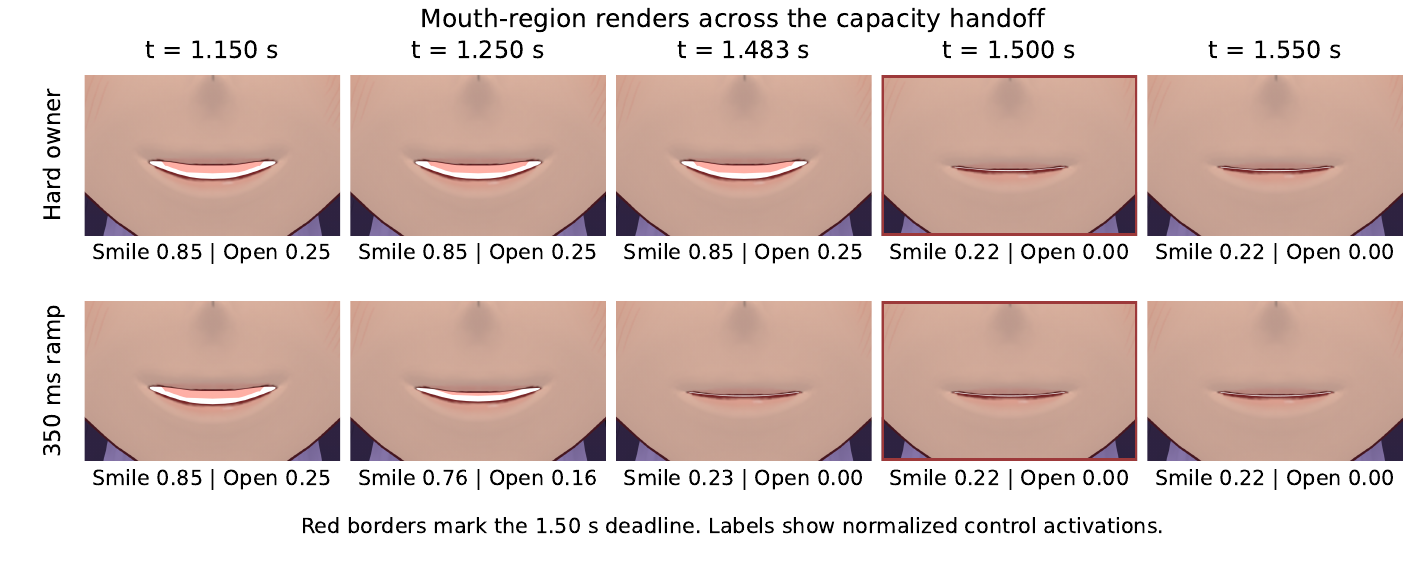}
\caption{Mouth-region renders before and across the reservation deadline. Every panel uses the same close-up camera; rows differ only in the ownership schedule. Smile and Open are normalized actuator activations. Red borders indicate the deadline. The earlier, gradual mouth change in the announced row reflects its advance reservation.}\label{fig:handoff_frames}
\end{figure*}

Figure~\ref{fig:handoff_trace} shows the trajectories. Figure~\ref{fig:handoff_frames} enlarges the mouth region at five fixed times spanning the advance transition and deadline. The announced condition receives advance notice and starts attenuating expression earlier; it trades advance expression change for continuity at the deadline. The owner is an abstract capacity schedule without synthesized speech.


\subsection{Reference-delay calibration and operating points}
The initial diagnostic settings have different response delays. We select a global EMA time constant using a common feasible step, minimizing the largest absolute $T_{90}$ difference across the three profiles and then its mean difference. Ties favor the larger time constant. The candidate set is
\begin{equation*}\begin{split}
\tau\in\{&0.030,0.045,0.055,0.065,0.070,\\
         &0.075,0.090,0.120,0.160\}\ \mathrm{s}.
\end{split}\end{equation*} The sensitivity sweep uses $\eta \in \{0.0125, 0.025, 0.050\}\text{ s}$. Selection uses the calibration $T_{90}$.

Dynamic test inputs are written before calibration and evaluated after the selected settings are locked. They comprise a joy--fear--anger interruption with different transition times, a new seeded jitter sequence, and a 60\% ownership reservation from 1.60 to 3.60 s following a 0.60 s expression onset. Each test lasts five seconds at 60 Hz. Three timed repetitions across twelve operating points and three profiles produce 324 dynamic trials and 97,200 controller samples, separate from the calibration probe and the initial diagnostic results in Table~\ref{tab:quantitative}. The split changes stimulus timing and noise between calibration and testing.

The calibration selects $\tau=0.070$ s for the principal EMA comparator, giving the same 150 ms reference delay as Full on every supplied profile. On the independent profiles, this comparator has both lower dynamic cue RMSE and lower jerk than Full. In the coupled configuration, Full has RMSE 0.1378 versus 0.1381 for the calibrated comparator, while its jerk is 3432 versus 2862 s$^{-3}$. These settings trade reconstruction error against jerk.\par
Outside the common scheduled capacity-decrease frame, Full respects the configured rate bounds, whereas the calibrated EMA can exceed them. Reference-delay matching therefore equalizes the specified step response without equalizing the whole trajectory or the constraint set. A separate rate-limited postfilter comparison would be needed to isolate joint temporal optimization from explicit rate enforcement.

\begin{table*}[!t]\centering\small
\caption{Reference-Delay-Calibrated Controller Comparison}\label{tab:delay_match}
\setlength{\tabcolsep}{5pt}
\renewcommand{\arraystretch}{1.2}
\begin{tabularx}{\linewidth}{@{}Xrrrrr@{}}\toprule
Method & $E$ & $J/10^3$ & Delay (ms) & CPU$_{95}$ & $R$\\\midrule
\multicolumn{6}{@{}l@{}}{\textit{Live2D / Hiyori}}\\[3pt]
Full & 0.1874 & 3.696 & 150.0 & 0.366 & 0\\
Static QP + EMA & 0.1866 & 2.232 & 150.0 & 0.450 & 0.1025\\
\midrule
\multicolumn{6}{@{}l@{}}{\textit{VRM / independent}}\\[3pt]
Full & 0.0759 & 4.773 & 150.0 & 0.444 & 0\\
Static QP + EMA & 0.0702 & 3.667 & 150.0 & 0.486 & 0.1131\\
\midrule
\multicolumn{6}{@{}l@{}}{\textit{VRM / composite mask}}\\[3pt]
Full & 0.1378 & 3.432 & 150.0 & 2.760 & 0\\
Static QP + EMA & 0.1381 & 2.862 & 150.0 & 5.044 & 0.0699\\
\bottomrule
\end{tabularx}
\par\smallskip
\begin{minipage}{\linewidth}\footnotesize
\textit{Note.} Three dynamic sequences are evaluated with three timed repetitions at 60 Hz. Full uses $\eta=0.025$ s and Static QP + EMA uses $\tau=0.070$ s. $E$: cue RMSE; $J$: RMS cue jerk (s$^{-3}$); delay: reference-step $T_{90}$ in milliseconds; CPU$_{95}$: per-frame allocation time (ms). $R$: largest activation change beyond the rate allowance, excluding the common budget-decrease frame at 1.60 s.
\end{minipage}
\end{table*}

Table~\ref{tab:delay_match} reports the principal matched setting. Its ordinary-frame rate diagnostic excludes only the common scheduled budget-decrease frame at 1.60 s for every method. Other input changes, the initial frame and capacity release remain included. This separates ordinary rate enforcement from the unavoidable hard-preemption case studied above.

\subsection{Coupling, convergence and projection cost}\label{sec:solver_cost}
We diagnose the matched-delay traces in Table~\ref{tab:delay_match} using their archived PGD iteration counts. A deterministic replay adds counters for the budget projections and reproduces every activation within $10^{-10}$; the original Windows CPU samples are retained, without substituting instrumented replay times. Each profile--method entry contains 2,700 frames from three dynamic tasks and three repetitions. Table~\ref{tab:solver_cost} reports the work performed.

\begin{table*}[!t]\centering\small
\caption{Solver Work in the Matched-Delay Traces}
\label{tab:solver_cost}
\begin{tabularx}{\linewidth}{@{}Xlrrrr@{}}\toprule
Profile & Method & PGD mean / 95th & Bisection frames & Binding calls/frame & $\kappa_2(H)$\\\midrule
Hiyori independent & Full & 1.56 / 2 & 11.3\% & 0.25 & 1.000\\
 & Static QP + EMA & 1.69 / 2 & 11.3\% & 0.30 & 1.000\\
VRM independent & Full & 1.61 / 2 & 24.7\% & 0.54 & 1.000\\
 & Static QP + EMA & 1.77 / 2 & 24.7\% & 0.57 & 1.000\\
VRM composite mask & Full & 9.00 / 18 & 91.2\% & 7.54 & 2.424\\
 & Static QP + EMA & 19.90 / 34 & 91.2\% & 14.60 & 8.707\\\bottomrule
\end{tabularx}
\par\smallskip
\begin{minipage}{\linewidth}\footnotesize
\textit{Note.} Each row covers 2,700 matched-delay frames. PGD iterations show mean / 95th percentile. Bisection frames contain at least one binding group projection; each binding call executes 45 threshold steps. $\kappa_2(H)$ is the spectral condition number of the corresponding solve. Projection counts include direct-map preparation, PGD initialization and updates, and final EMA reprojection when used. CPU timings in Table~\ref{tab:delay_match} retain the archived Windows measurements. Deterministic replay validates activations and solver-work counts.
\end{minipage}
\end{table*}

For the independent VRM profile, Full requires 1.61 iterations on average (95th percentile: 2), versus 9.00 (18) for the coupled profile. The static comparator increases from 1.77 (2) to 19.90 (34). At least one 45-step budget bisection is triggered in 24.7\% of independent-VRM frames and 91.2\% of coupled frames. In the coupled case, Full performs 7.54 binding group projections per frame on average, compared with 14.60 for Static QP + EMA. Repeated projections inside the longer PGD solves explain why their cost grows more than the one-time twelve-variable matrix operations.

The Hessian is $H=C_m^\top C_m+(\lambda+\eta/\Delta t)I$ for Full and $H_0=C_m^\top C_m+\lambda I$ for the static solve. Both independent profiles have unit condition number. For the coupled profile, the off-diagonal Gram-matrix Frobenius norm is 1.610, and the measured spectral condition numbers are $\kappa_2(H)=2.424$ and $\kappa_2(H_0)=8.707$. The temporal diagonal shift therefore improves conditioning as well as encoding continuity. Non-diagonal structure alone does not imply worse conditioning; the measured spectrum and active projections establish the numerical mechanism in this case. No matched-delay frame reaches the 60-iteration cap.

Full's archived CPU$_{95}$ increases from 0.444 to 2.760 ms (6.22-fold); the coupled static comparator takes 5.044 ms. Full's 2.760 ms uses 16.6\% of a 60 Hz frame budget and 24.8\% of a 90 Hz budget (16.67 and 11.11 ms). The largest observed coupled Full sample is 4.757 ms, compared with 8.623 ms for the static comparator. Percentile and observed maximum timings are empirical allocation costs, not worst-case execution-time guarantees. Rendering, inference and behavior processing require the remaining frame budget; an integrated 90 Hz application has not been measured here.

\begin{figure*}[!t]\centering
\includegraphics[width=\linewidth]{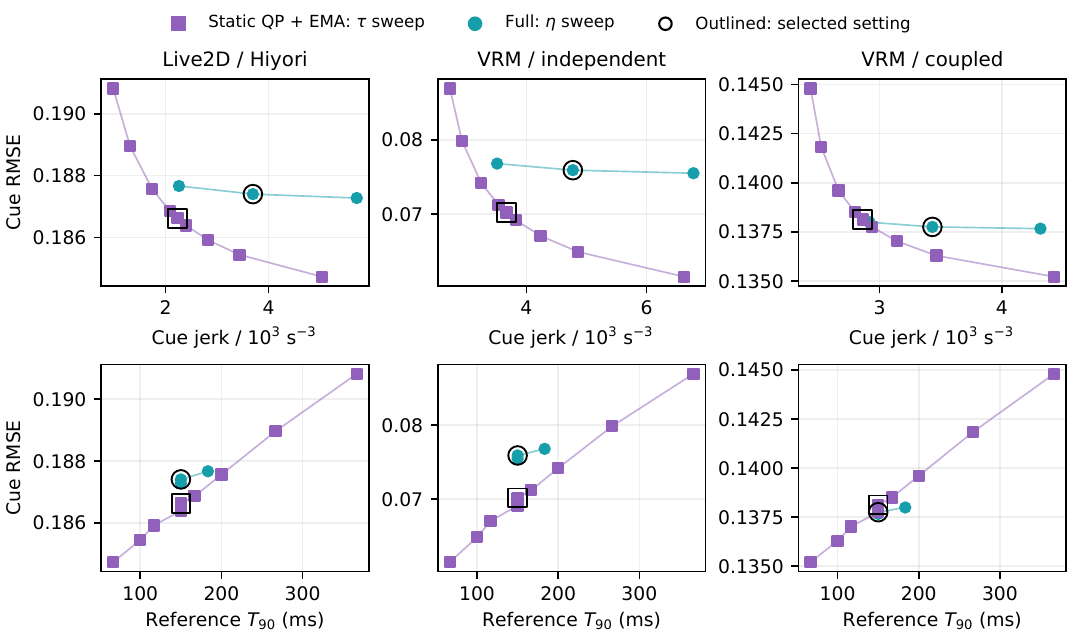}
\caption{Operating points on the separate dynamic inputs. Top: reconstruction error versus cue jerk. Bottom: reconstruction error versus the reference calibration delay. Cyan circles show the Full temporal-weight sweep in $\eta$ (eta); purple squares show the Static QP + EMA time-constant sweep in $\tau$ (tau). Each column holds one profile fixed. Points are enumerated controller settings, not independent perceptual observations. Outlined points mark Full at $\eta = 0.025$~s and its selected EMA comparator at $\tau = 0.070$~s. Rate compliance is reported separately in Table~\ref{tab:delay_match}.}\label{fig:operating_points}
\end{figure*}

\section{Perceptual evaluation}
\subsection{Design and participants}
We conducted the evaluation through an institutional online platform using a $2$ (control method) $\times\,8$ (avatar model) block-randomized between-participant design. The analyzed sample comprised 386 participants, balanced across 16 cells with 24--25 participants per cell. Each method had 193 participants. All cells contained 24 participants except the two VRM A cells, which contained 25. Each participant was assigned one method and one avatar and evaluated ten distinct base stimuli sampled without replacement from the stimulus pool. Stimulus selection was randomized. The platform used anonymous participant identifiers and de-identified demographic information by design before export. Participants provided digital informed consent before the assessment.

\subsection{Avatars and stimulus generation}
The eight avatars comprise four Live2D models and four VRM models, reported as anonymized condition codes L2D A--D and VRM A--D. These codes preserve confidentiality for licensed, closed-source character assets. The set includes the publicly available Hiyori Momose Live2D sample\cite{hiyoriasset} and the AvatarSample\_B VRM sample\cite{vroidasset} illustrated in the controlled cases. Third-party character binaries and Cubism Core remain subject to their respective license terms.

We generated the perceptual stimuli at 60 Hz. Full used $\eta=0.025$ s and $\lambda=0.002$, and Static QP + EMA used $\tau=0.070$ s with the same reconstruction regularization. The calibrated reference-step response was $T_{90}=150$ ms for both methods. The 500 base stimuli comprised 250 state tasks and 250 transition tasks, covering seven intended emotion categories and all 42 directed transitions between different categories. Each base stimulus specifies an expressive target and timing; its controller and avatar are determined by the participant's assigned condition.

State clips lasted five seconds, with the final second defining the recognition window. Transition clips lasted six seconds: the start window was 1--2 s, the transition began at 2 s and lasted 0.4, 0.8 or 1.2 s, and the end window was 5--6 s. This structure separates identification of the two endpoints from evaluation of the movement between them.

\subsection{Tasks and measures}
For a state clip, participants selected the emotion shown in the final window and rated whole-clip naturalness. For a transition clip, they separately selected the start and end emotions and rated naturalness, transition smoothness and change clarity. The categories were neutral, joy, sadness, anger, fear, surprise and calm, with an additional uncertain option. The three motion-quality ratings used seven-point scales, with larger values indicating greater naturalness, continuity or clarity.

The completed sample contained 1,950 state responses and 1,910 transition responses. State recognition requires the observed final label to match the authored target. Transition start and end recognition compare their respective windows; joint transition recognition requires both to match. The pooled endpoint measure therefore includes one judgment per state response and two per transition response. An uncertain emotion response counts as unmatched. Naturalness is evaluated over all clips; smoothness and clarity are evaluated over transition clips.

\subsection{Statistical analysis}
We analyzed all completed records from the 386 participants. Correctness was recomputed from the authored target labels and observed responses, and participant- and stimulus-level totals were reconciled. Group differences are Full minus Static QP + EMA. Recognition differences are expressed in percentage points; rating differences retain their seven-point scale units.

We estimated 95\% percentile intervals with 10,000 crossed participant--stimulus bootstrap draws, using seed 20261028. Participants were resampled within method-by-model strata, retaining all ten responses from each selected participant. Base-stimulus IDs were independently resampled across the shared pool, and each response received the product of its participant and stimulus multiplicities\cite{owen2012bootstrap}. The two endpoint judgments of a transition retained the same weight. Model conditions were held fixed. Participant-only intervals are included as a sensitivity analysis in the supplementary results. Reported intervals are marginal intervals for the individual outcomes.

\subsection{Recognition and motion quality}
\begin{figure*}[!t]\centering
\includegraphics[clip,width=\linewidth]{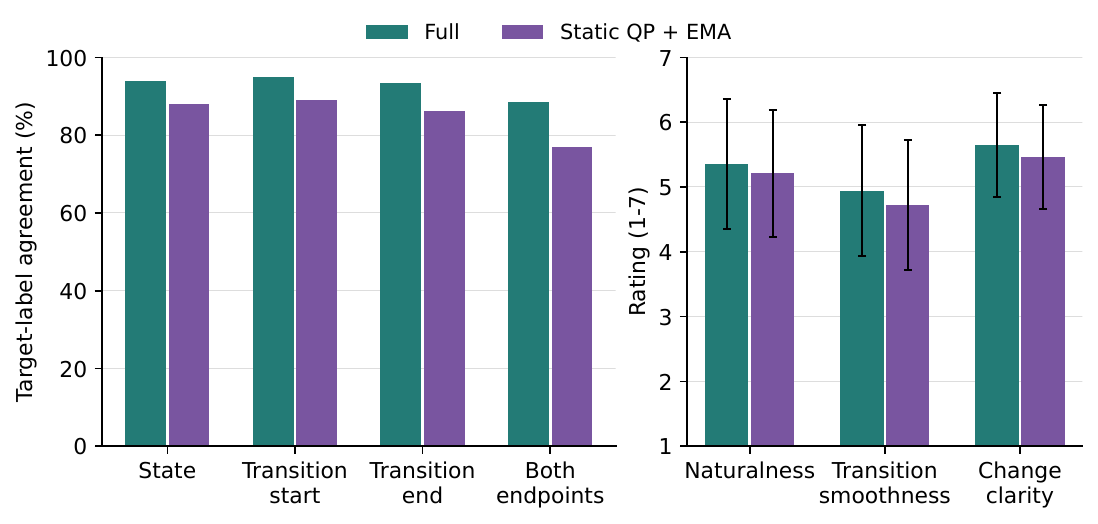}
\caption{Recognition and motion-quality outcomes at a matched 150 ms reference-step response. Recognition bars separate the state task and transition recognition measures. Rating whiskers show response-level standard deviations. Group-difference intervals are reported in Table~\ref{tab:human_outcomes}.}\label{fig:perceptual_outcomes}
\end{figure*}

Full matched 2,724 of 2,894 endpoint judgments (94.13\%), compared with 2,525 of 2,876 (87.80\%) for Static QP + EMA. The difference was 6.33 percentage points (95\% interval 3.81 to 8.88). Table~\ref{tab:human_outcomes} and Figure~\ref{fig:perceptual_outcomes} report the recognition endpoints and motion-quality ratings.

\begin{table*}[!t]\centering\small
\caption{Perceptual Outcomes at the Matched 150 ms Reference Response}\label{tab:human_outcomes}
\setlength{\tabcolsep}{4pt}
\begin{tabularx}{\linewidth}{>{\raggedright\arraybackslash}Xrrr}
\toprule Outcome & Full & Static QP + EMA & Difference [95\% interval]\\\midrule
All endpoint judgments & 94.13\% & 87.80\% & +6.33 {[}3.81, 8.88{]}\\
State recognition & 94.00\% & 88.01\% & +5.99 {[}1.78, 10.22{]}\\
Transition start & 95.02\% & 89.11\% & +5.91 {[}1.87, 10.22{]}\\
Transition end & 93.36\% & 86.26\% & +7.10 {[}2.42, 12.00{]}\\
Both transition endpoints & 88.49\% & 76.85\% & +11.64 {[}6.14, 17.47{]}\\
Naturalness (1--7) & 5.36 $\pm$ 1.00 & 5.21 $\pm$ 0.98 & +0.146 {[}-0.001, 0.292{]}\\
Transition smoothness (1--7) & 4.94 $\pm$ 1.01 & 4.72 $\pm$ 1.00 & +0.217 {[}0.031, 0.398{]}\\
Change clarity (1--7) & 5.65 $\pm$ 0.80 & 5.46 $\pm$ 0.80 & +0.190 {[}0.059, 0.315{]}\\
\bottomrule\end{tabularx}

\par\smallskip
\begin{minipage}{\linewidth}\footnotesize
\textit{Note.} Participants: 193 / 193; responses: 1,930 / 1,930 (Full / comparator). Percentages pool applicable categorical judgments. Rating entries show response-level mean \ensuremath{\pm} SD. Differences are Full minus Static QP + EMA, in percentage points for recognition and scale points for ratings. Brackets give marginal 95\% crossed participant--stimulus bootstrap intervals (10,000 draws). State responses: 966 / 984; transition responses: 964 / 946; all endpoint judgments: 2,894 / 2,876 (Full / comparator).
\end{minipage}
\end{table*}

Both transition endpoints were recognized in 88.49\% of Full responses and 76.85\% of comparator responses, an increase of 11.64 percentage points (95\% interval 6.14 to 17.47). Change clarity increased by 0.190 points (95\% interval 0.059 to 0.315), and transition smoothness increased by 0.217 points (95\% interval 0.031 to 0.398). Mean naturalness was 5.36 versus 5.21; its 0.146-point difference had an interval from -0.001 to 0.292, including zero.

\begin{figure*}[!t]\centering
\includegraphics[clip,width=\linewidth]{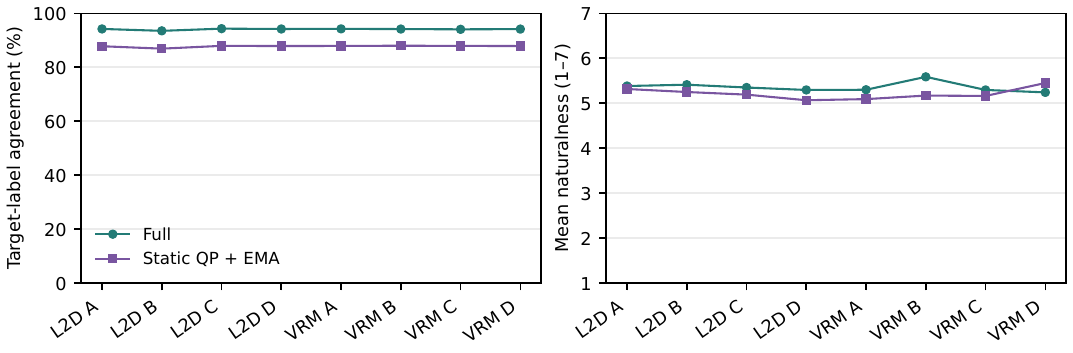}
\caption{Recognition and naturalness across the eight anonymized avatar conditions. Each method has 24 participants per condition, except VRM A with 25. L2D A--D and VRM A--D preserve the identities of licensed character assets.}\label{fig:perceptual_models}
\end{figure*}

Figure~\ref{fig:perceptual_models} reports descriptive recognition and naturalness by anonymized avatar condition. The eight labels identify the licensed model conditions used in the randomized design.

\begin{figure*}[!t]\centering
\includegraphics[clip,width=\linewidth]{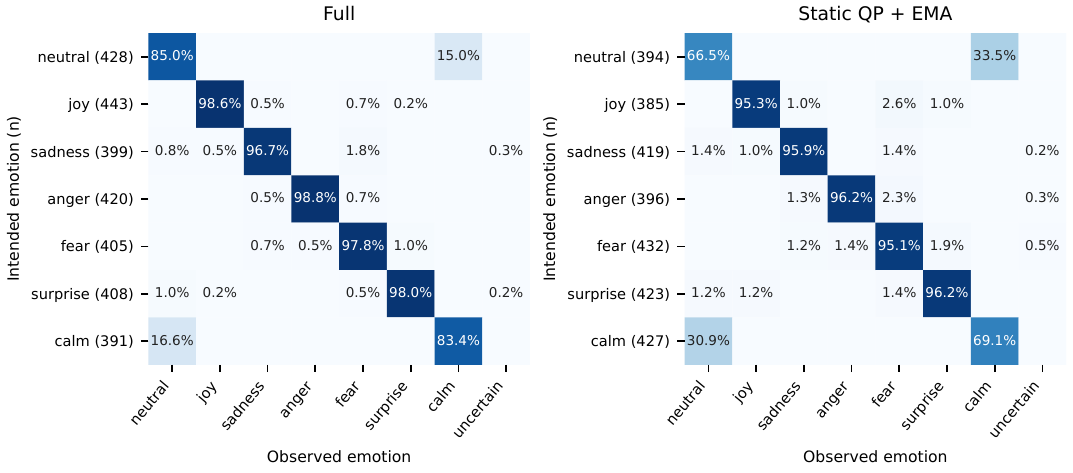}
\caption{Row-normalized intended-versus-observed emotion judgments, including both endpoints of transition responses. Row labels show judgment counts; uncertain responses remain in the denominator, and blank cells indicate zero observations. The neutral--calm cross-confusion rate is 32.16\% for Static QP + EMA and 15.75\% for Full when the two target categories are combined.}\label{fig:perceptual_confusion}
\end{figure*}

Figure~\ref{fig:perceptual_confusion} shows the corresponding intended-versus-observed emotion distributions. Neutral was confused with calm in 33.5\% of baseline judgments and 15.0\% of Full judgments. The reverse confusion decreased from 30.9\% to 16.6\%. Combining the two target categories, cross-confusion fell from 32.16\% (264/821) to 15.75\% (129/819), a 51.0\% relative reduction.

\section{Discussion}
\subsection{Perceptual separability beyond aggregate tracking error}
The central perceptual result is improved separability of intended affect at a matched response delay. Full increased pooled endpoint agreement by 6.33 percentage points and joint transition recognition by 11.64 percentage points. The confusion matrices reveal where much of this difference occurs: neutral-to-calm confusion decreased from 33.5\% to 15.0\%, and calm-to-neutral confusion decreased from 30.9\% to 16.6\%. Combining the two target categories gives a 51.0\% relative reduction in cross-confusion.

These distinctions are easy to obscure in an aggregate error over twelve proxy coordinates. A weak cue can contribute little to average RMSE while changing whether an observer reads a face as neutral or deliberately relaxed. The pattern is consistent with joint temporal allocation preserving subtle expressive cues during the onset and replacement of an expression. The increase in change clarity, 0.190 scale points with a positive interval from 0.059 to 0.315, supports this interpretation. It also explains why similar proxy-space errors can coexist with different recognition outcomes.

Recognizing both endpoints is a stricter transition measure than recognizing either endpoint alone. Its improvement from 76.85\% to 88.49\% indicates that the benefit extends to communicating a complete change of affect. Naturalness provides a different assessment: its mean difference was only 0.146 points and its interval included zero. The results therefore support clearer and more recognizable expression without requiring a large improvement in overall naturalness.

\subsection{What joint temporal allocation contributes}
The formulation's distinct behavior arises from the interaction of temporal tracking, coupled cue effects and active constraints. In the special case $C_m^\top C_m=I$, with no active bounds or budgets, Equation~\ref{eq:objective} reduces to
\begin{equation}
\begin{aligned}
\mathbf{q}_t&=\alpha\frac{C_m^\top\mathbf{z}_t}{1+\lambda}+(1-\alpha)\mathbf{q}_{t-1},\\
\alpha&=\frac{1+\lambda}{1+\lambda+\eta/\Delta t}.
\end{aligned}
\label{eq:ema_limit}
\end{equation}
This limiting case is a first-order filter of the regularized static solution. On a coupled rig, however, a single actuator may affect several expressive cues, and its feasible range can change with another behavior's ownership. Joint optimization coordinates that compromise with the preceding activation. The coupled profile has nonzero Gram-matrix off-diagonals (Table~\ref{tab:solver_cost}), so the identity-map reduction does not apply. The benchmark shows that this coordination does not uniformly minimize both RMSE and jerk; the perceptual study identifies the resulting benefit in expression interpretation at the matched operating point.

The capacity-handoff case illustrates a complementary scheduling tradeoff. Advance notice allows the controller to reduce an expression over several frames before the ownership deadline, replacing an abrupt mouth change with a gradual transition. Immediate hard ownership still has priority when a new budget conflicts with the rate allowance. Thus continuity depends on both the allocation rule and the information available to the behavior scheduler.

\subsection{Scope and extensions}
The perceptual comparison tests the complete controller at a matched delay; it does not independently manipulate coupling strength, temporal regularization and active budgets. The geometric diagnostic isolates a concrete coupled-control mechanism, while the participant results establish the between-method perceptual contrast. A coupling-by-method experiment would test whether coupling mediates that contrast.

The evaluation covers four Live2D and four VRM conditions, including publicly available samples and licensed character models. Anonymized condition codes retain the latter assets' confidentiality. The cue-effect model remains a local additive approximation: antagonistic deformations, saturation and nonlinear expression combinations motivate richer calibration or nonlinear extensions. The current solver handles disjoint resource groups; overlapping ownership envelopes require a more general projection. Further work can connect the allocator to synchronized speech, physics and native actions and measure adaptation effort and request-to-visible latency in a production runtime.

\section{Conclusion}
SoulLink realizes continuous affect through capability-aware allocation over heterogeneous Live2D and VRM controls. Its cue-effect model, joint temporal objective and ownership-dependent budgets connect expressive targets to feasible actuator motion. Controlled cases expose the influence of coupled controls and show how an announced handoff spreads a capacity transition over time. In the 386-participant block-randomized study, Full increased intended-emotion agreement by 6.33 percentage points and joint transition recognition by 11.64 percentage points at a matched 150 ms reference response. Combined neutral--calm confusion was reduced by 51.0\%, and change clarity improved by 0.190 scale points. The naturalness difference remained small. The findings support joint temporal allocation as a practical way to preserve subtle expression distinctions and make emotion transitions more recognizable across the evaluated avatar conditions.

\begin{samepage}
\section*{Declarations}
\subsection*{Code and data availability} Source snapshots, allocation code, controlled simulation traces, rendered illustrations and the anonymized perceptual-study dataset, together with the bootstrap analysis scripts, accompany this manuscript. The public project repositories provide the Live2D backend at \url{https://github.com/nanlingyin/soullink-emotion-sdk} and the VRM backend at \url{https://github.com/nanlingyin/SoullinkEmotion3D}. The anonymized perceptual data and analysis are publicly available at \url{https://github.com/nanlingyin/SoullinkEmotion3D/tree/paper/paper}. Third-party character binaries and Live2D Cubism Core are accessed separately under their respective license terms.
\par\end{samepage}
\subsection*{Human participants} The perceptual evaluation was conducted through an institutional online evaluation platform in accordance with local ethical guidelines and the ethical principles of the Declaration of Helsinki\cite{wma2024}. Under the institution's provisions for minimal-risk educational and perceptual software evaluation, this study of anonymous subjective ratings qualified for ethical exemption. The stimuli were non-sensitive computer-generated avatar animations, and no personally identifiable or biometric data were collected. All participants provided digital informed consent before beginning the assessment.
\subsection*{AI-assisted preparation} OpenAI ChatGPT/Codex assisted with manuscript preparation and parts of the research implementation.
\subsection*{Asset attribution} Hiyori Momose is a Live2D Inc. sample character\cite{hiyoriasset}. AvatarSample\_B is a pixiv VRoid sample\cite{vroidasset}. Rendered illustrations retain the original character designs.

\balance
\bibliographystyle{IEEEtran}\bibliography{references,references_recent}
\end{document}